\documentclass{optica-article}

\journal{opticajournal} 

\articletype{Research Article}

\usepackage{hyperref}
\usepackage{lineno}
\usepackage{xcolor}

\begin{document}

\title{Comparing continuous and pulsed nitrogen-vacancy DC magnetometry in the optical-power-limited regime}

\author{Maggie Wang,\authormark{1} Michael Caouette-Mansour,\authormark{1} Adrian Solyom,\authormark{1} Lilian Childress,\authormark{1,*}}

\address{\authormark{1}McGill University, 3600 Rue University, Montreal QC H3A2T8}

\email{\authormark{*}lilian.childress@mcgill.ca} 


\begin{abstract*} 
Ensembles of nitrogen-vacancy (NV) center spins in diamond offer a robust, precise and accurate magnetic sensor. As their applications move beyond the laboratory, practical considerations including size, complexity, and power consumption become important. Here, we compare two commonly-employed NV magnetometry techniques -- continuous-wave (CW) vs pulsed magnetic resonance -- in a scenario limited by total available optical power. We develop a consistent theoretical model for the magnetic sensitivity of each protocol that incorporates NV photophysics - in particular, including the incomplete spin polarization associated with limited optical power; after comparing the models' behaviour to experiments, we use them to predict the relative DC sensitivity of CW versus pulsed operation for an optical-power-limited, shot-noise-limited NV ensemble magnetometer. We find a $\sim 2-3 \times$ gain in sensitivity for pulsed operation, which is significantly smaller than seen in power-unlimited, single-NV experiments~\cite{dreau_avoiding_2011}. Our results provide a resource for practical sensor development, informing protocol choice and identifying optimal operation regimes when optical power is constrained. 


\end{abstract*}

\section{Introduction}
The nitrogen-vacancy (NV) center in diamond has become a sensor of choice for a wide range of applications and techniques, from detection of nuclear magnetic resonance with small analyte volumes ~\cite{mamin_nanoscale_2013, staudacher_nuclear_2013} to imaging magnetic textures in novel materials~\cite{casola_probing_2018}.  Its long-lived ground state spin can be prepared and detected optically, even at room temperature, permitting detection of any environmental variable that perturbs the spin states, including electric~\cite{dolde_electric-field_2011} or magnetic fields~\cite{rondin_magnetometry_2014}, temperature~\cite{neumann_high-precision_2013, kucsko_nanometre-scale_2013}, and strain~\cite{trusheim_wide-field_2016}. Notably, the Zeeman splitting of the NV spin sublevels has led to widespread applications in magnetometry.

Optically detected magnetic resonance (ODMR) is the simplest modality for NV sensing of DC or low-frequency magnetic fields. Under optical illumination, the NV polarizes preferentially into the $m_s = 0$ sublevel of its ground state spin triplet ($S = 1$); resonant microwave (MW) driving of its $m_s = 0$ to $m_s = \pm 1$ transitions induces a reduction in its fluorescence, allowing optical detection of the spin transition frequencies, which in turn reveal the magnetic field components along and perpendicular to the NV symmetry axis. Full vector magnetometry can be realized by identifying transition frequencies for all four possible orientations of the NV center within the diamond crystal lattice~\cite{pham_magnetic_2011}. ODMR can be performed with continuous-wave (CW) optical and MW excitation or in a pulsed manner, by alternating optical excitation with MW $\pi$ pulses that flip the spin state when resonant. Despite its simplicity, ODMR remains a competitive practical  magnetometry approach, offering high dynamic range with sensitivities approaching more sophisticated Ramsey-based techniques in some cases~\cite{dreau_avoiding_2011, barry_sensitivity_2020}, without the need for phase estimation algorithms~\cite{waldherr_high-dynamic-range_2012, nusran_high-dynamic-range_2012}. 

While CW ODMR is particularly straightforward to implement, pulsed ODMR outperforms CW ODMR by eliminating optical broadening of the spin resonances, with an order of magnitude improvement in sensitivity observed in early single-NV studies~\cite{dreau_avoiding_2011}. Nevertheless, additional considerations apply when translating these results to ensemble magnetometers, which offer improved sensitivity by addressing a larger sample volume. In particular, an important practical limitation for ensemble magnetometers is the power budget, especially the optical power that can be supplied to a small-form-factor device. At the same time, sensors designed for biological studies may need to limit optical intensity to avoid damaging samples.  Limited optical power affects not only the fluorescence signal strength, but also the rate of optical spin polarization, and an inhomogeneous distribution of optical power induces spatially-varying dynamics of the NV spin polarization. 

Here, we revisit the question of the sensitivity gains that can be expected from pulsed vs CW ODMR for an ensemble magnetometer with limited optical power, operating in the optical-shot-noise-limited regime. While recent studies have examined aspects of this comparison in different contexts~\cite{zhang_pulsed_2022, poulsen_investigation_2021, zhang_diamond_2021}, we focus on the following question: given limited optical power but the ability to optimize other experimental controls, how much gain can be expected upon upgrading an ensemble magnetometer from CW to pulsed operation? We develop a consistent model of the NV photophysics to describe both CW and pulsed protocols; after validating its behavior against experimental data, we use the model to predict the sensitivity gain associated with pulsed ODMR both for single NVs and for ensembles. Aside from imposing fixed optical power, we allow each protocol to find its optimal operation point in microwave power and pulse timing, ultimately finding that in most situations pulsed ODMR only improves the sensitivity for power-limited ensemble magnetometry by a factor of 2 to 3. 
We also elucidate the optimal regimes of operation for CW and pulsed ODMR, providing a resource for practical sensor development, particularly in applications where the cost, weight, or energy budget imposes constraints on optical power and device complexity. 

\section{Modelling ODMR signals from a single NV}
The magnetic field sensitivity of each ODMR scheme is determined by the NV fluorescence signal as the MW drive is swept in frequency across the spin resonance condition. The greater the depth and the narrower the linewidth of the fluorescence feature, the more precisely the magnetic field can be measured. 
Specifically, the sensitivity $\eta$ depends on the uncertainty in the inferred magnetic field $\Delta B$ after measurement for duration $T$ according to $\eta = \Delta B \sqrt{T}$. For an ODMR signal with Lorentzian lineshape (appropriate to CW signals), probed at its point of maximum slope, the optical-shot-noise-limited sensitivity to fields along the NV axis is
\begin{equation}
\label{eq:etaCW}
\eta_{CW}= \frac{2 \Delta\nu }{3 c \gamma_{NV}} \sqrt{\frac{\frac{4}{3}-c}{F_0}},
\end{equation}
where $\gamma_{NV} = 2.8$ MHz/G is the NV gyromagnetic ratio, $\Delta \nu$ is the FWHM linewidth (in frequency), $c$ is the fluorescence contrast, and $F_0$ is the off-resonance fluorescence rate. Similarly, a pulsed ODMR signal with a lineshape approximated by a Gaussian would have, at its point of maximum slope, 
\begin{equation}
\label{eq:etapulsed}
\eta_{pulsed} = \frac{\Delta \nu}{2 c \gamma_{NV}}\sqrt{\frac{\sqrt{e}\left(\sqrt{e} -c\right)}{F^0_{avg}\log{4}}},
\end{equation}
where $F^0_{avg}$ is the average fluorescence rate (averaged over the total time for the pulse sequence) for off-resonant microwaves. In the low-contrast limit, Eq.~\ref{eq:etaCW}, \ref{eq:etapulsed} reduce to standard expressions~\cite{barry_sensitivity_2020, dreau_avoiding_2011}.

The optical-shot-noise-limited magnetic sensitivity can thus be predicted from a model that finds the NV fluorescence rate as a function of microwave detuning from resonance, and extracts the off-resonant fluorescence rate, linewidth, and contrast. In this section, we consider the signals from a single NV center, neglecting hyperfine structure for the time being (as appropriate e.g. for magnetic fields near the excited state level anti-crossing~\cite{jacques_dynamic_2009}), and assuming that the external magnetic field is sufficiently large that only one of the two magnetic-dipole-allowed transitions (e.g. $m_s = 0$ to $m_s = +1$) is driven by near-resonant microwaves. 

\subsection{NV photophysics}
We model the fluorescence of an NV center using rate equations and/or optical Bloch equations, building on previous work ranging from simple and intuitive two-level models examining only the driven Zeeman states~\cite{dreau_avoiding_2011, zhang_detection_2021} to models incorporating the ground and excited triplet as well as singlet states~\cite{el-ella_optimised_2017, poulsen_investigation_2021}. Importantly, we seek to use the same model (with the same parameters) for both CW and pulsed regimes, such that we can directly compare outcomes, and we include effects of finite spin polarization. For the room-temperature, low-optical-power scenario we consider, we employ an effective 4-level model that offers a balance between relative simplicity and accuracy in modeling the system dynamics.  

\begin{figure}[ht!]
\centering\includegraphics[width=7cm]{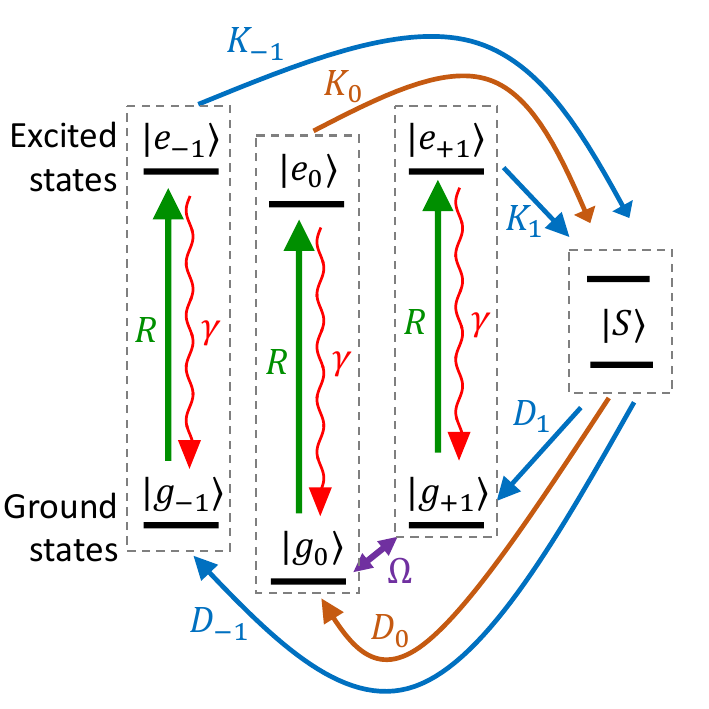}
\caption{The rate equation model used incorporates optical excitation at rate $R$, radiative relaxation at rate $\gamma$, relaxation into the singlets from $|e_i\rangle$ at rate $K_i$, and relaxation from the singlets into $|g_i\rangle$ at rate $D_i$. Microwaves drive transitions between $|g_0\rangle$ and $|g_1\rangle$ at Rabi frequency $\Omega$. The dashed gray boxes indicate the populations considered in the 4-level model: the total population in each triplet spin projection and the total singlet population.}
\label{fig1}
\end{figure}

We first consider evolution in the absence of MW driving. At low optical excitation powers, ionization rates (which scale quadratically with optical power~\cite{aslam_photo-induced_2013}) are negligible, so the relevant states of the negatively charged NV center (see Fig.~\ref{fig1}) comprise three ground state spin sublevels $\{|g_0\rangle, |g_{-1}\rangle, |g_{+1}\rangle\}$, three excited state spin sublevels $\{|e_0\rangle, |e_{-1}\rangle, |e_{+1}\rangle\}$, and a singlet state $|S\rangle$ (neglecting the $\sim$GHz relaxation between the two singlets~\cite{doherty_nitrogen-vacancy_2013}). We consider incoherent transition rates between the states as illustrated in Fig~\ref{fig1} and a coherent MW drive that is near resonance with one spin transition, e.g. $m_s = 0$ to $m_s = +1$. 
Note that we neglect intrinsic relaxation between the spin sublevels, as it is typically much slower than all other rates in the system for NV centers at room temperature or below~\cite{jarmola_temperature-_2012}. 

To simplify the model, we assume that the ground and excited states instantaneously reach their steady-state proportions within each spin projection, effectively neglecting any dynamics on timescales similar to or faster than the rate of equilibration on the optical transition. Specifically, we assume that the population in $|e_{i}\rangle$ is $e_i(t) = R g_i(t)/(\gamma +K_i)$, where $g_i(t)$ is the instantaneous ground state population in $|g_i\rangle$, $R$ is the spin-conserving optical excitation rate,  $K_i$ is the rate of decay out of $|e_i\rangle$ into $|S\rangle$, $\gamma$ is the radiative relaxation rate, and $i \in  \{-1, 0, 1\}$. Perhaps surprisingly, this assumption works well across a wide range of values of $R$: for $R\ll \gamma$, equilibration happens on a timescale set by $\gamma$, but the dynamics of interest occur on a much longer timescale set by $R$; for $R \gg \gamma$, $R$ dominates the equilibration rate, while the dynamics of interest are limited by the much slower deshelving rate $D_0 + 2 D_1$. In each case the ground-excited state equilibration occurs much faster than the dynamics of interest.  

We can thereby model the total population in each spin state $m_i(t) = e_i(t) + g_i(t)$ via the following rate equations:

\begin{equation}
\label{eq:photophysics}
    \begin{array}{clcr}
    m_i'(t) &=& -K_i \dfrac{R}{R+K_i + \gamma}m_i(t) + D_i S(t)\\
    \\[-12pt]
    S'(t) &=& \displaystyle \sum_{i \in \{-1, 0, 1\}} \left(K_i \dfrac{R}{R+K_i + \gamma}m_i(t) -D_i S(t)\right)
    \end{array},
\end{equation}
where $D_i$ is the rate of decay out of $|S\rangle$ into $|g_i\rangle$ and $S(t)$ is the instantaneous population in $|S\rangle$. While it is possible to further simplify the model by eliminating the singlet state and combining the $m_s= \pm 1$ states, the results only approximate the full dynamics for extremely weak optical powers such that $R\ll D_i$, which is too restrictive for our desired application. 

Throughout this work, we consider fixed values for the radiative and nonradiative decay rates, given by the average values measured in reference~\cite{gupta_efficient_2016}: $\gamma = 66.50$ MHz, $K_0 = 10.78$ MHz, $K_{-1} = K_{1} = 91.07$ MHz, $D_0 = 4.835$ MHz, and $D_{-1} = D_{1} = 1.063$ MHz. 

\subsection{CW ODMR}

To add in coherent MW excitation on the $m_s = 0 \leftrightarrow m_s = 1$ transition, we modify the equations of motion to include the coherence $\rho_{01}$ of this transition~\cite{el-ella_optimised_2017}:
\begin{equation}
\begin{array}{rcl}
\rho_{01}'(t) &=& - (\Gamma_2 + i \delta) \rho_{01}(t) + i  \dfrac{\Omega}{2} \left(m_{1}(t)-m_0(t)\right)\\
\\[-12pt]
m_{1}'(t) &=& -i \dfrac{\Omega}{2}\left(\rho_{10}(t) - \rho_{01}(t)\right) + D_{1}S(t) - \dfrac{K_{1}R}{R +K_1+ \gamma}m_{1}(t)\\
\\[-12pt]
m_0'(t) &=& i \dfrac{\Omega}{2}\left(\rho_{10}(t) - \rho_{01}(t)\right)  + D_{0}S(t) - \dfrac{K_{0}R}{R +K_0+ \gamma}m_{0}(t),
\end{array}
\end{equation}
where $\Omega$ is the MW Rabi frequency (angular frequency), $\delta$ is the microwave detuning from resonance (angular frequency), $\rho_{10} = \rho_{01}^*$, and the equations of motion for $S(t)$ and $m_{-1}(t)$ remain as they were in Eq.~\ref{eq:photophysics}. The dephasing rate $\Gamma_2$ depends both on the intrinsic dephasing time $T_2^*$ of the spin and on the optically-induced dephasing rate, which we approximate as $\Gamma_2 = R + 2\sqrt{\log{2}}/T_2^*$ (this expression uses the relationship between dephasing rate and $T_2^*$ appropriate for quasistatic noise~\cite{de_sousa_electron_2009} and assumes that the spin dephases as soon as it is optically excited). 
Treating intrinsic spin dephasing in this manner is an approximation, as many
diamond samples exhibit slowly-evolving magnetic noise; while convolving results
with a distribution of detunings would better approximate such noise processes,
the added model complexity would significantly reduce its utility. Thus, for the
sake of simplicity, we treat the spin dephasing solely within a white noise model.
Note that these equations of motion presume that the majority of the spin population is in the ground state, where it can be driven by near-resonant MW, corresponding to $R\ll \gamma$. Fortunately we will see that this is also the regime in which CW ODMR works best. 

Solving the optical Bloch equations in the steady state allows us to calculate the fluorescence rate  $F(\delta) = \epsilon \gamma \sum_i \left(\dfrac{R}{R+K_i+\gamma}\right)m_i^{SS} + b R$, where $\epsilon$ is the collection efficiency of the experiment, $m_i^{SS}$ are the steady-state populations of the spin sublevels, and the adjustable parameter $b$ accounts for background fluorescence from other sources, which typically scales linearly with excitation intensity. We highlight the detuning dependence, as we use $F(\delta)$ to determine the linewidth $\Delta \nu$ (FWHM) and contrast $c$ of the dip in fluorescence that occurs at resonant MW driving, as well as the off-resonant fluorescence rate $F_0$. Substitution into Eq.~\ref{eq:etaCW} yields the sensitivity of the CW ODMR protocol.

\begin{figure}[ht!]
\centering\includegraphics[width=10cm]{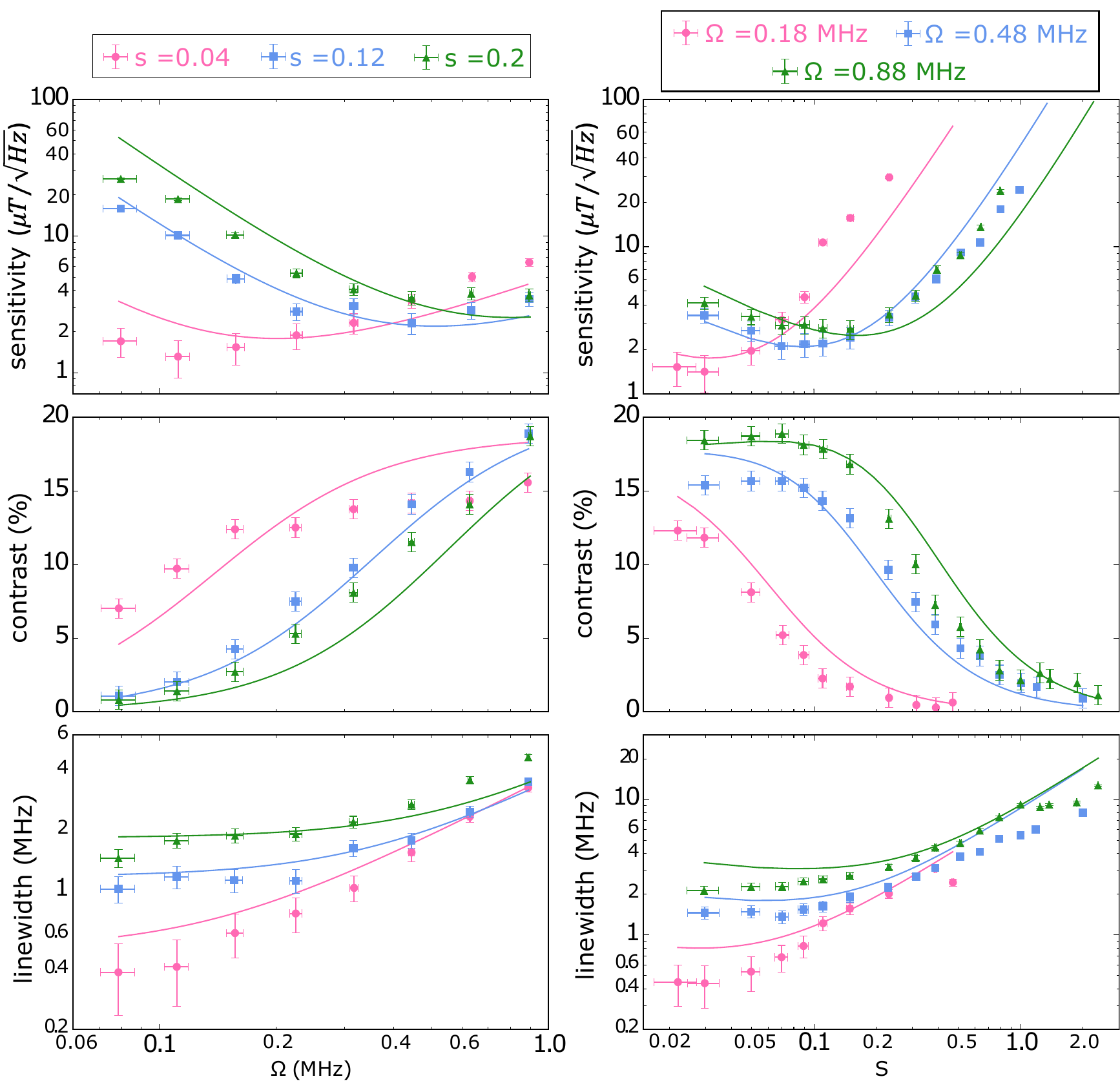}
\caption{Comparison of CW model predictions to published data from ref.~\cite{dreau_avoiding_2011} (reprinted with permission from \href{https://doi.org/10.1103/PhysRevB.84.195204}{Dreau \textit{et al.}, Phys. Rev. B 84, 195204 (2011)}), showing the sensitivity, contrast and linewidth as a function of Rabi frequency (left) and saturation parameter (right). Legends at top apply to all figures in the column. We use the experimentally measured $T_2^* = 3\upmu$s and fit adjustable parameters of collection efficiency $\eta$ = 0.98\% and background fluorescence $b = 0.0031$  (81 kcts/s background at $s = 1$).}
\label{fig:dreauCW}
\end{figure}

To validate our model, we compare its predictions to data published in reference~\cite{dreau_avoiding_2011} (see Fig.~\ref{fig:dreauCW}), examining linewidth, contrast, and sensitivity as a function of optical excitation rate. We use the measured value of $T_2^* = 3\upmu$s, with other rate parameters as given above. 
To convert from the underlying optical excitation rate $R$ in our model to the saturation parameter $s$ used experimentally, we use the background-subtracted off-resonant fluorescence rate to extract $R_{sat} = \dfrac{(2 D_1 + D_0)K_0 K_1 + (2 D_1 K_0 + D_0 K_1)\gamma}{2D_1 K_0 + D_0 K_1 + K_0 K_1}$, with $s = R/R_{sat}$. 
This conversion is approximate, as the model neglects ionization, which can distort the experimental saturation curve from the theoretical shape. However for optical powers well below a saturation intensity, the distortion of the saturation curve by ionization is minimal.
Despite this issue, our model qualitatively describes the data; although the agreement is not exact, we emphasize that this is a fixed-parameter model, with only the collection efficiency $\epsilon$ and background fluorescence $b$ as adjustable variables, and it accurately captures the trends observed experimentally. 

\begin{figure}[ht!]
\centering\includegraphics[width=12cm]{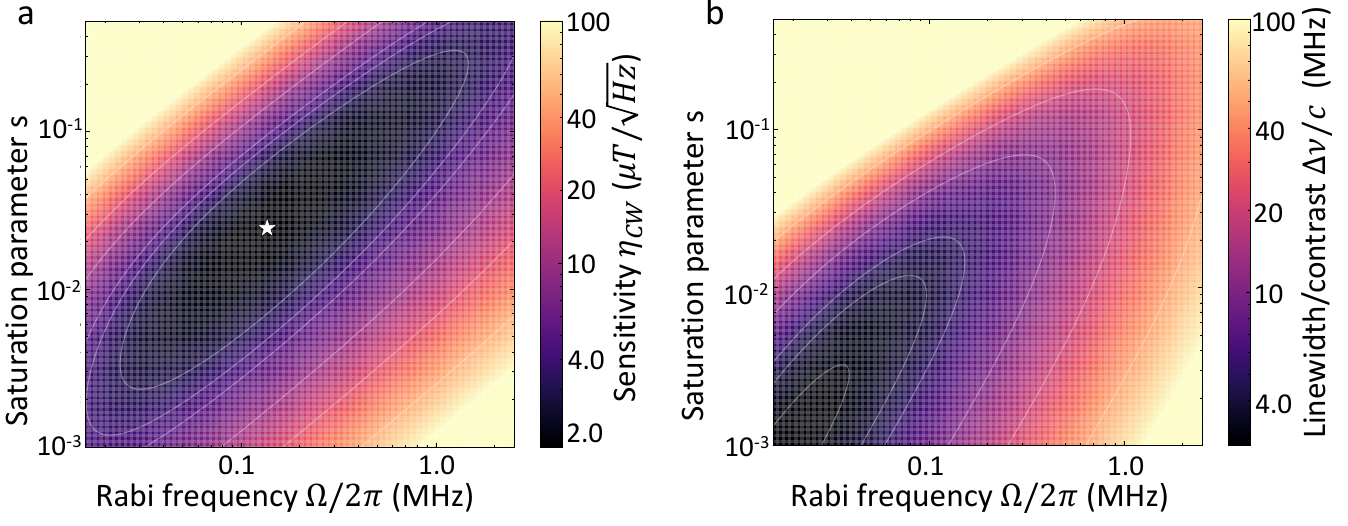}
\caption{Figures of merit for the CW protocol as a function of Rabi frequency and saturation parameter, calculated for the same parameters as Fig. 2 ($T_s^*=3\upmu$s, $\epsilon=0.98\%$ and $b=0.0031$). (a) Modeled sensitivity. Star indicates point of optimal sensitivity. (b) 
Within the limitations of our
model, the calculated linewidth/contrast exhibits no optimal point.} 
\label{fig:etaCW}
\end{figure}

Using the collection efficiency and background values that match the experimental data from Fig.~\ref{fig:dreauCW}, we calculate the sensitivity as a function of Rabi frequency and saturation parameter as shown in Fig.~\ref{fig:etaCW}a. The optimal sensitivity of  1.69 $\upmu$T/$\sqrt{\text{Hz}}$ occurs for $\Omega = (2\pi) 0.136$ MHz and $s = $ 0.024. 
We also show a commonly-cited figure of merit, the ratio of the linewidth and contrast (see Fig. 3b). Because our CW model neglects intrinsic longitudinal spin relaxation and quasistatic spin dephasing mechanisms, this ratio is minimized for vanishingly small optical and microwave power, illustrating that the optimal operation point in Fig. 3a does not arise from minimizing linewidth/contrast. If spin relaxation and/or quasi-static dephasing mechanisms were added to the model, the linewidth/contrast would exhibit a minimum value at finite $s$ and $\Omega$. However, for typical parameter values, this minimum occurs at such small $s$ and $\Omega$ that the qualitative conclusion remains the same: the optimal operation point observed in Fig. 3a arises not directly from optimizing linewidth/contrast but rather from competition between minimizing linewidth/contrast (which prefers weaker optical and MW excitation) and generating sufficient signal $F_0$, which requires more power.

\subsection{Pulsed ODMR}
We now seek to employ the same photophysics to model the performance of pulsed ODMR. Pulsed ODMR alternates a MW pulse of duration $t_\pi$ with an optical pulse of duration $t_L$, with a short wait interval $t_w$ after the optical pulse and before the MW to allow for relaxation from the singlets. Fluorescence counts are recorded during the initial time period $\tau$ of the laser pulse (any remaining time during the laser pulse serves to repolarize the spin towards $m_s=0$). Typically, this pulse sequence is repeated many times at each detuning, so we examine the fluorescence counts observed after a large number of cycled pulses (the "pulsed steady-state"), which we cast as a matrix diagonalization problem as described below. This approach allows us to incorporate the effects of incomplete spin repolarization during the optical pulse.

First, we consider evolution under MW driving in the absence of optical excitation -- this will determine three parameters needed for finding the sensitivity: the MW pulse duration $t_\pi$, the linewidth $\Delta \nu$ and the probability that the MW pulse flipped the spin $c_\pi$. Unlike in CW ODMR, where the spin dephasing is strongly influenced by near-Markovian optical processes, for pulsed ODMR nuclear spin noise dominates; we therefore treat the electron spin decoherence with a quasi-static classical noise model~\cite{de_sousa_electron_2009}. We assume that the spin resonance frequency is constant during each microwave pulse, but follows a Gaussian probability distribution over the course of an experiment. For example, Rabi oscillations would be modeled by convolving the probability of a spin flip under driving with angular Rabi frequency $\Omega$ at angular detuning $\delta$ for time $t$ with a Gaussian probability distribution of angular detunings with variance $\Delta^2 = 2/(T_2^*)^2$ 
and zero mean. We use the result to calculate that the pulse duration $t_\pi$ that maximizes the probability of a spin flip; note that due to dephasing $t_\pi$ is not always equal to $\pi/\Omega$. 

Once $t_\pi$ is identified, we find the lineshape for spin resonance probed with a pulse of duration $t_\pi$. Specifically, we convolve the spin-flip probability for an ideal two-level system driven by a pulse of duration $t_\pi$ and detuning $\delta$ with a Gaussian distribution of detunings with mean $\delta_0$ and variance $\Delta^2 = 2/(T_2^*)^2$. The resulting lineshape (as a function of angular frequency $\delta_0$) reveals the FWHM linewidth in frequency $\Delta \nu$ and the probability $c_\pi$ of a spin flip on resonance. 
While we use fully numerical values in what follows, the linewidth is well approximated by $\Delta \nu \approx \sqrt{0.0646326~ \Omega^2 + \dfrac{4 \log{2}}{\pi^2 (T_2^*)^2}}$ with $\Omega$ given as an angular frequency and $\Delta \nu$ as a frequency, found by combining in quadrature the limiting values for large and small dephasing. Furthermore, it is worth noting that both the approximate and numerically determined linewidths neglect interplay between the MW pulse and optical pulse, which could be a fruitful avenue for future study.

To calculate the sensitivity via Eq.~\ref{eq:etapulsed}, we still need to find the fluorescence contrast $c$ and the average fluorescence rate $F_{avg}^0$ in the the pulsed steady state, which can be determined by modeling the photophysics of the NV center with the MW $\pi$ pulse on and off resonance. We represent the NV center with a 4-dimensional vector of populations, $\mathbf{p} = \{m_{-1}(t), m_{0}(t), m_1(t), S(t)\}$, and find 4-dimensional matrix representations of its evolution during each time period of the pulse pattern. During the optical pulse, by representing Eqs.~\ref{eq:photophysics} as a matrix equation $\mathbf{p}'(t) = \mathbf{\Lambda} ~\mathbf{p}(t)$, we can find the population evolution as $\mathbf{p}(t) = e^{\mathbf{\Lambda} t}\mathbf{p}(0)$. We model relaxation during the wait time as the matrix $\mathbf{W} = \lim_{t \rightarrow \infty} e^{\mathbf{\Lambda} t}$ with $R = 0$ (we assume that the wait time is sufficiently long to fully relax the singlets to avoid the complication of singlet relaxation during the microwave pulse).   Finally, we model the resonant MW pulse by 
\begin{equation} 
\label{eq:Pi}
\mathbf{\Pi} = \left(\begin{array}{cccc}
1&0&0&0\\
0&1-c_\pi& c_\pi&0\\
0&c_{\pi}&1-c_\pi&0\\
0&0&0&1\end{array}\right),
\end{equation}
and assume that the off-resonant MW pulse acts as the identity matrix. Over the course of a single pulse sequence consisting of a laser pulse, wait time, and MW pulse, the populations therefore evolve as
\begin{equation}
    \mathbf{p}_\text{final} = \mathbf{\Pi}\mathbf{W}e^{\mathbf{\Lambda}t_L}\mathbf{p}_{\text{initial}}.
\end{equation}
The pulsed steady state (defined just before the laser pulse) corresponds to the condition $\mathbf{p}_\text{final} = \mathbf{p}_\text{initial}$, i.e. to populations that are an eigenvector of $\mathbf{\Pi}\mathbf{W}e^{\mathbf{\Lambda}t_L}$ with eigenvalue 1. 

We solve two eigenvalue problems, one for the eigenvector $\mathbf{p}_1$ corresponding to $\mathbf{p_\text{final} }= \mathbf{p_\text{initial}}$ in by Eq.~\ref{eq:Pi}, and the other for $\mathbf{p}_0$, following the same procedure with $\mathbf{\Pi}$ replaced by the identity matrix. From this, we can compute the on-resonance (off-resonance) fluorescence counts during time $\tau$ by finding $\epsilon \gamma$ times the integrated excited state population:  

$$C_{1(0)} = \epsilon \gamma \int_0^\tau  \left\{\frac{R}{R+K_{-1}+\gamma}, \frac{R}{R+K_{0}+\gamma},\frac{R}{R+K_{1}+\gamma}, 0\right\}.(e^{\mathbf{\Lambda}t}\mathbf{p}_{1(0)})dt.$$ The contrast is then given by $c = 1-C_1/C_0$ and the off-resonant count rate is given by $F^0_{avg}=C_0/(t_{\pi}+t_w+t_L)$. 
Combining $c$ and $F_{avg}^0$ with the linewidth $\Delta \nu$ determined above, we can calculate the pulsed ODMR sensitivity in Eq.~\ref{eq:etapulsed} as a function of $R$ (which we convert to $s$ as described above), $\Omega$, $\tau$, and $t_L$. 

\begin{figure}[ht!]
\centering\includegraphics[width=10cm]{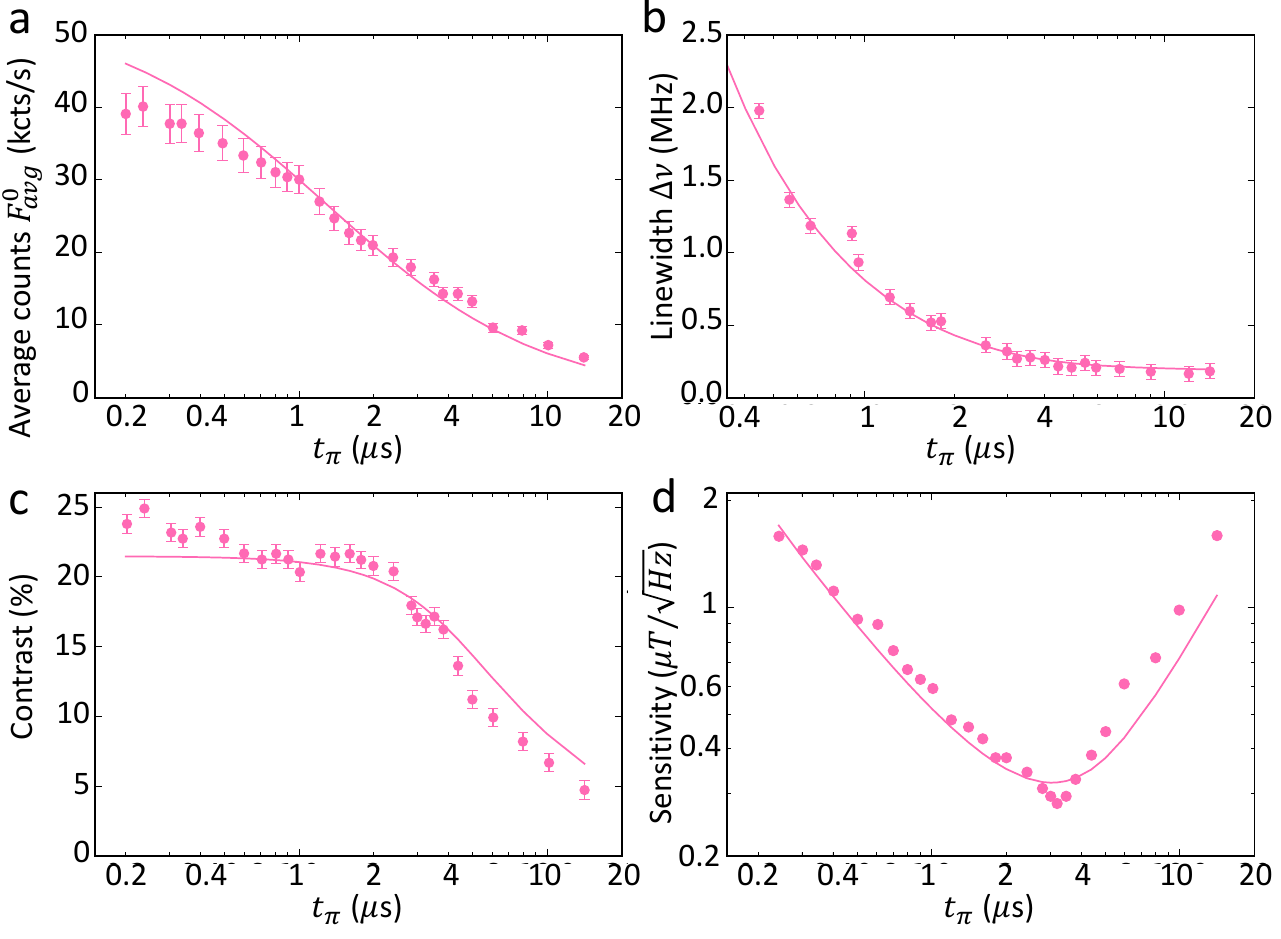}
\caption{Model comparison to pulsed ODMR experiment for a single NV. Points are experimental data from reference~\cite{dreau_avoiding_2011} (reprinted with permission from \href{https://doi.org/10.1103/PhysRevB.84.195204}{Dreau \textit{et al.}, Phys. Rev. B 84, 195204 (2011)}), solid lines are model predictions using fixed parameters $T_2^* = 3\upmu$s $s = 1.2$, $t_w = 1 \upmu$s, $\tau = 300 ns$, $t_L = 300 ns$, $\epsilon = 0.98\%$, $b = 0.0031$. (a) Average fluorescence count rate $F_{avg}^0$. (b) Linewidth $\Delta \nu$. (c) Contrast $c$. (d) Sensitivity. 
}
\label{fig:dreaupulsed}
\end{figure}

\begin{figure}[ht!]
\centering
\includegraphics[width=10cm]{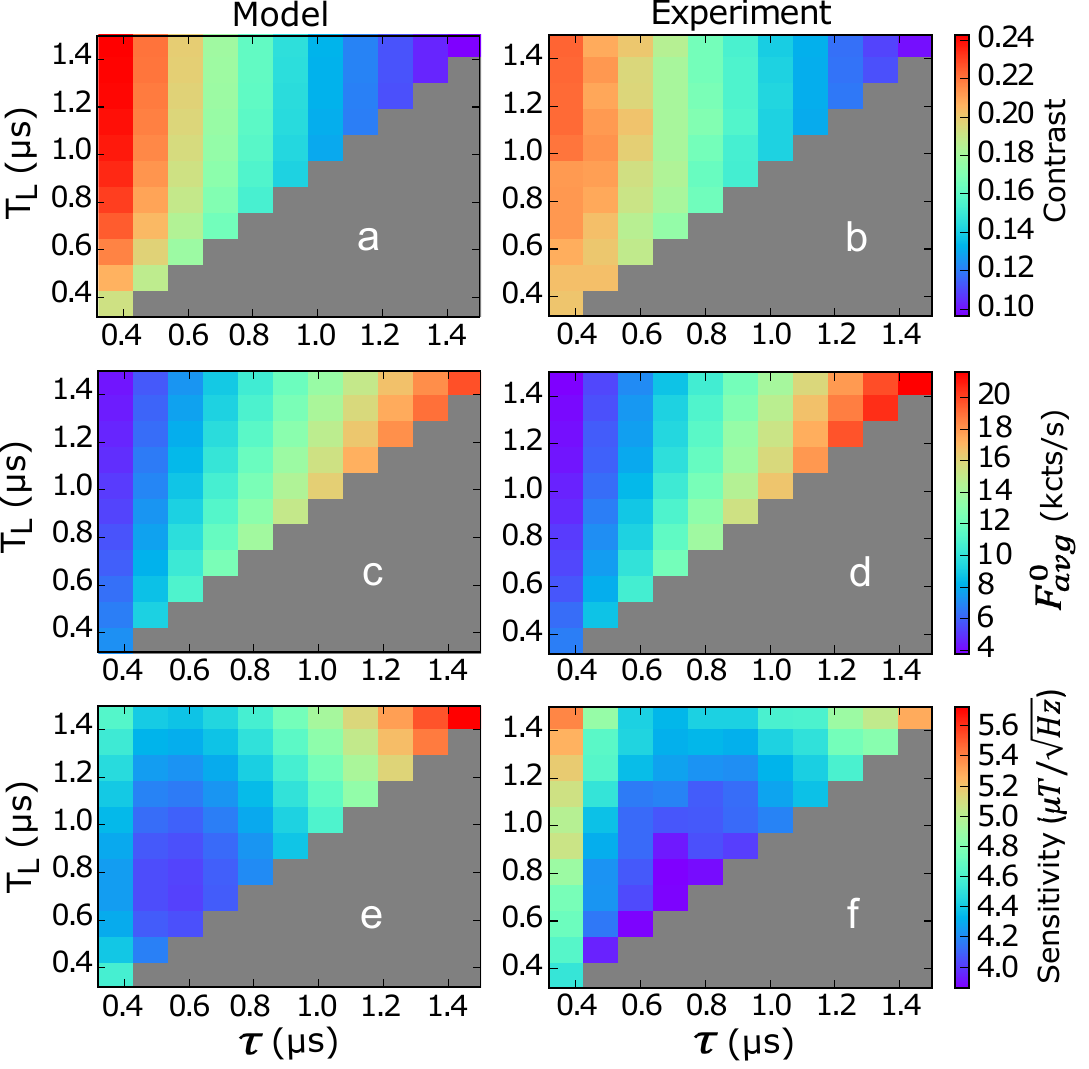}
\caption{Dependence on polarization and readout time. Left-hand column shows values modeled using $T_2^* = 2.3~\upmu$s, $s = 0.3$ ($T_2^*$ and $s$ are measured independently), $t_\pi = 0.208~\upmu$s, $t_w = 1~\upmu$s ($t_\pi$ and $t_w$ correspond to experimental values), $\epsilon = 0.43\%$, and $b = 0.0017$ ($\epsilon$ and $b$ are fit to experimental data). (a)-(b) Contrast. (c)-(d) Average fluorescence rate. (e)-(f)  Sensitivity.  
}
\label{fig:Michael}
\end{figure}

We again seek to validate our model in comparison to experimental data. Figure~\ref{fig:dreaupulsed} shows how the modelled linewidth, contrast, counts and sensitivity vary as a function of $t_\pi$, as compared to data from reference~\cite{dreau_avoiding_2011}. We use the same values of collection efficiency $\eta=0.98\%$ and background $b = 0.0031$ as were found from the CW data (see Fig.~\ref{fig:dreauCW}), and use the experimentally measured or given values of $T_2^* = 3\upmu$s, $s = 1.2$, $t_w = 1 \upmu$s, $\tau = 300$ ns, and $t_L = 300$ ns for our model parameters, while $\Omega$ is inferred from $t_\pi$. To examine the behaviour of our model as a function of polarization and readout time, which was not available in previously published data sets, we performed pulsed ODMR experiments on a single NV center. Figure~\ref{fig:Michael} shows the modeled vs measured contrast, average fluorescence rate, and sensitivity as a function of pulse parameters (see caption for details). Again, given that our model allows adjustment of only the collection efficiency and background, we observe a reasonable level of agreement.  

\begin{figure}[ht!]
\centering\includegraphics[width=7cm]{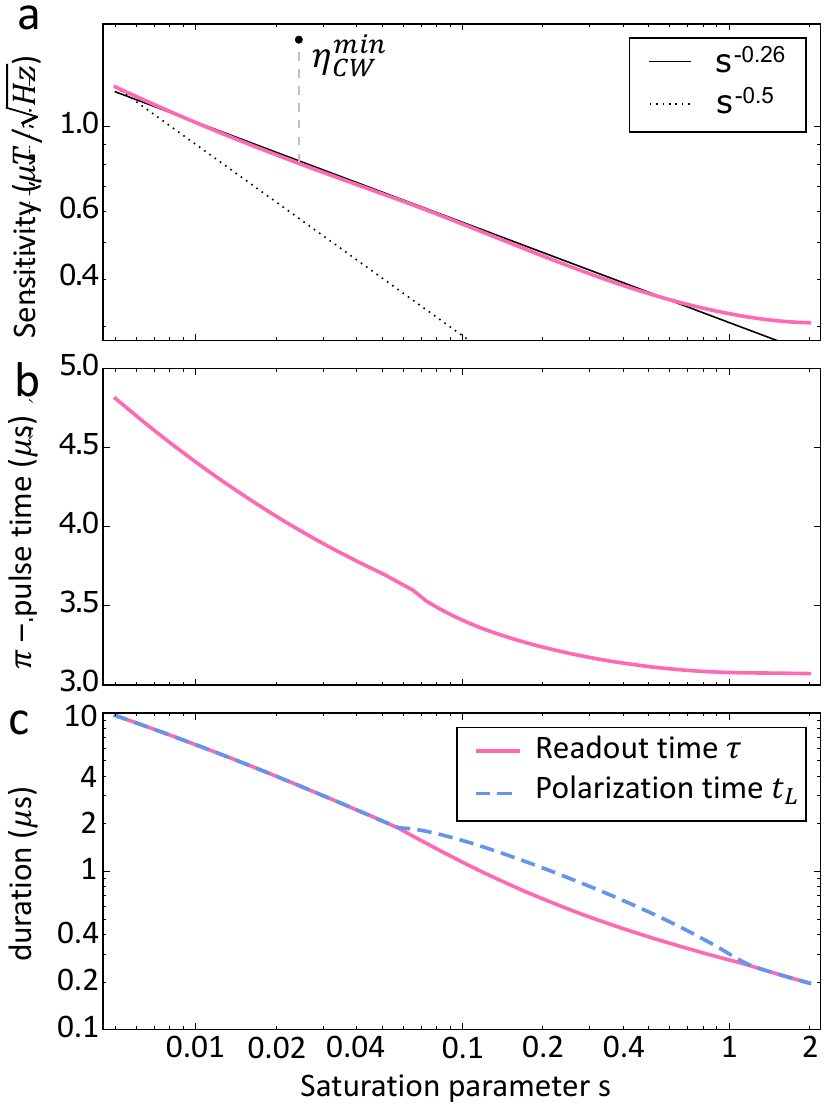}
\caption{Optimal sensitivity and operating conditions for pulsed ODMR on a single NV, as a function of saturation parameter. In all cases, model parameters are $T_2^* = 3\upmu$s, $t_w = 1 \upmu$s, $\epsilon = 0.98\%$, $b = 0.0031$. 
(a) Optimal sensitivity as a function of saturation parameter, optimized over $\tau, T_L,$ and $\Omega$. The optimal CW sensitivity (point) and associated saturation parameter (dashed line) are shown for comparison. Straight lines illustrate $s^{-0.26}$ and $s^{-0.5}$ power law dependences. (b) Duration of the $\pi$-pulse 
 $t_\pi$ that yields optimal sensitivity (note that $\Omega$ and $T_2^*$ determine $t_\pi$). (c) Polarization time $t_L$ and readout time $\tau$ that optimize sensitivity. 
}
\label{fig:etapulsed}
\end{figure}

Having verified that our model provides a good approximation of NV dynamics, we can use it to explore operating regimes for pulsed-ODMR magnetometry. Because the pulsed-ODMR sensitivity depends on the pulse parameters as well as the Rabi frequency and optical power, it offers a richer optimization landscape than CW ODMR. In particular, we find that the optimal pulse sequence varies significantly with the available optical power.  Fig.~\ref{fig:etapulsed} shows how the optimal sensitivity and operating conditions vary with optical power, using $T_2^* = 3 \upmu$s, $t_w = 1 \upmu$s, $\eta = 0.98\%$, $b = 0.0031$ as for the experimental data from reference~\cite{dreau_avoiding_2011}. Unsurprisingly, the optimal durations for polarization $t_L$ and readout $\tau$ decrease with optical power; for low optical power they are equal to each other, as the benefit of extending $t_L$ to obtain slightly increased spin polarization is outweighed by the increase in time required to achieve it.  
 $\tau  = \tau_L$ is also optimal at high power, matching the regime chosen experimentally in reference~\cite{dreau_avoiding_2011}. While the bifurcation in optimal pulse timings at moderate optical powers may appear surprising, the benefit of having $\tau \neq \tau_L$ is minimal (here it improves the sensitivity by at most $7$ nT$/\sqrt{Hz}$). This result points to the possibility of simplifying a device by accumulating photon counts without gating detectors. 

 The optimal $\pi$ pulse duration also decreases as the optical power increases; this behaviour arises from two effects: (1) the dependence of the sensitivity on the duration of the pulse pattern $\eta \propto \sqrt{T}$ and (2) the dependence of the spin state preparation (in the pulsed steady state) on the MW spin flip probability. The first mechanism is straightforward: at high optical powers, $t_\pi$ represents the majority of $T$, so reducing $t_\pi$ directly benefits the sensitivity, whereas at low optical powers $T$ is dominated by $T_L$, and reductions in $t_\pi$ have minimal effect. The second mechanism is more subtle: because we are examining the pulsed steady state, the spin state preparation fidelity depends on both the probability that the optical pulse repolarizes the spin and the probability that the MW pulse flips it. As an extreme example, with no MW pulse, the pulsed steady state is completely polarized, whereas with a MW spin flip, the state preparation will be imperfect if the optical pulse cannot fully repolarize the spin. As a result, for low optical powers where optical polarization is incomplete, the initial spin state preparation is better when the MW pulse is less effective, giving a slight benefit to longer $t_\pi$ values. These two mechanisms have the same tendency to promote longer $t_\pi$ times for lower optical powers, leading to the observed trend.
 With these optimized parameters, the sensitivity decreases with optical power, approximately following a $s^{1/4}$ power law. 

Overall, using a consistent model for both protocols, the sensitivity obtained for pulsed ODMR is better than for CW ODMR (see Fig.~\ref{fig:etapulsed}a). However, it is worth noting that if we constrain the optical power for pulsed ODMR to be equal to the optimal power for CW ODMR, the improvement is substantially reduced: at the optimal CW ODMR saturation parameter $s = 0.024$, optimized pulsed ODMR yields a sensitivity of 802 nT/$\sqrt{\text{Hz}}$, only 2.11 times better than the optimal CW ODMR sensitivity. 

\section{Ensemble ODMR}
With an understanding of how pulsed and CW ODMR of a single NV center behave as a function of optical power, we can extend our model to ensembles of NV centers, examining magnetometer sensitivity when the total optical power is constrained. 

For an ensemble magnetometer, there is typically a distribution of optical intensities across the sample. We assume an optical excitation beam with fixed total power and a spatial mode approximated by a cylindrical Gaussian intensity profile with variable waist $\sigma$ (1/$e^2$ intensity); we also assume a spatially varying collection efficiency for the emitted fluorescence with the same cylindrical Gaussian shape and a maximum value of 1\%. In our model, this beam illuminates a diamond similar to varieties available commercially off the shelf, with a NV concentration of 300 ppb, $T_2^* = 1\upmu$s, and a thickness of 500 $\upmu$m. We convert the excitation beam intensity to a distribution of saturation parameters using a saturation intensity of 1.1 mW/$\upmu$m$^2$~\cite{rittweger_far-field_2009}. Using the models developed above, we integrate the total fluorescence from a uniform distribution of NVs throughout the volume (with the integration truncated at a radius of 10 $\sigma$ for efficiency); we assume that only one quarter of the NVs are resonant with the MW drive, while the other three quarters correspond to orientations that are always off-resonant with the MW drive, and fluoresce at the same off-resonant rate (neglecting any decrease in fluorescence arising from off-axis magnetic fields). 
Additionally, we assume the polarization of the excitation light excites all 4 orientations equally, reducing the 
saturation parameter $s$ by a factor of 2/3 for all orientations. Since we are most interested in the optical power limitations, we assume a homogeneous Rabi frequency, though in practice this consideration may limit available sample volume. Ultimately, the integrated fluorescence signal allows us to extract the contrast, linewidth, and off-resonant count rate needed to estimate the sensitivity. 



\begin{figure}[ht!]
\centering\includegraphics[width=7cm]{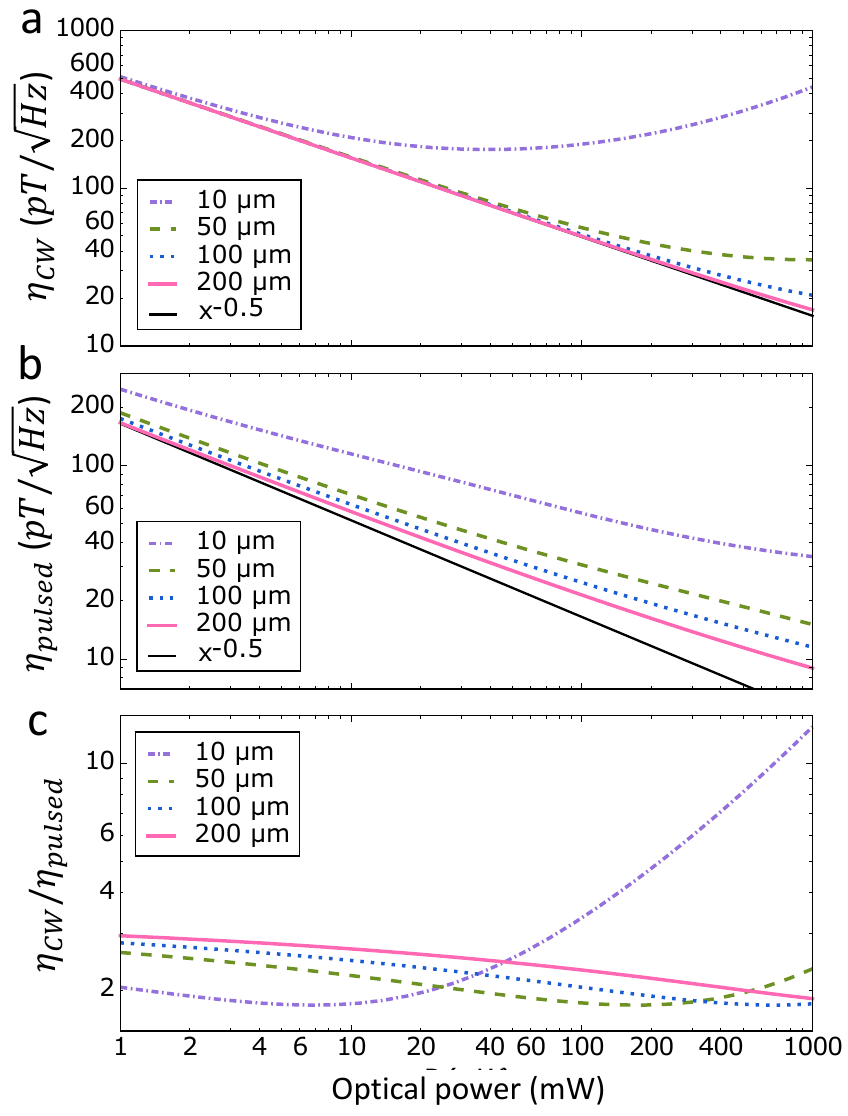}
\caption{Comparison of sensitivity for NV ensembles employing CW vs pulsed protocols, as a function of total optical power ($T_2^* = 1 \upmu$s, NV density = 300 ppb). (a) CW sensitivity vs power for an optical beam radius of 10 $\upmu$m, 50 $\upmu$m, 100 $\upmu$m, and 200 $\upmu$m.   (b) Pulsed sensitivity vs power at the same beam waists. A straight line with a slope of -1/2 is added for comparison in both (a) and (b).  (c) Ratio of CW to pulsed sensitivity for the same set of beam waists. }
\label{fig:ensemblepulsed}
\end{figure}

Figure ~\ref{fig:ensemblepulsed} shows the CW and pulsed ODMR sensitivity as a function of total optical power, for beam waists ranging from 10 to 200 $\upmu$m. Both schemes prefer the largest beam waist attainable, due to the saturable nature of the NV transition: a greater fluorescence signal can be obtained by spreading the excitation power over a larger number of NV centers. Even for CW ODMR, where one might anticipate a preferred beam waist related to the optimal single-NV intensity, lower optical intensity is preferred:  
because a decrease in optical intensity occurs in conjunction with an increase in beam area, both the linewidth-to-contrast ratio and (unlike for a single NV) the total fluorescence signal improve as the beam waist increases. In practice, however, the sample size will be limited by Rabi frequency homogeneity and diamond dimensions, so we limit our analysis to beam waists of $200 \upmu$m or less.

In Fig.~\ref{fig:ensemblepulsed}a, the minimum CW ODMR sensitivity closely follows a $s^{-1/2}$ power law, which further indicates that the improvement in sensitivity can be largely attributed to increases in the fluorescence count rate. Nevertheless, for each beam waist there is an optimal power, above which optical spin dephasing overwhelms increased fluorescence, thereby worsening sensitivity. In contrast, the pulsed ODMR sensitivity improves slightly less with optical power (see Fig.~\ref{fig:ensemblepulsed}b), but does not exhibit an optimal power. 
 Ultimately, as shown in Fig.~\ref{fig:ensemblepulsed}c, 
for larger beam waists where CW works well, pulsed ODMR only improves the sensitivity by a factor of 2 to 3. 

\begin{figure}[ht!]
\centering\includegraphics[width=6cm]{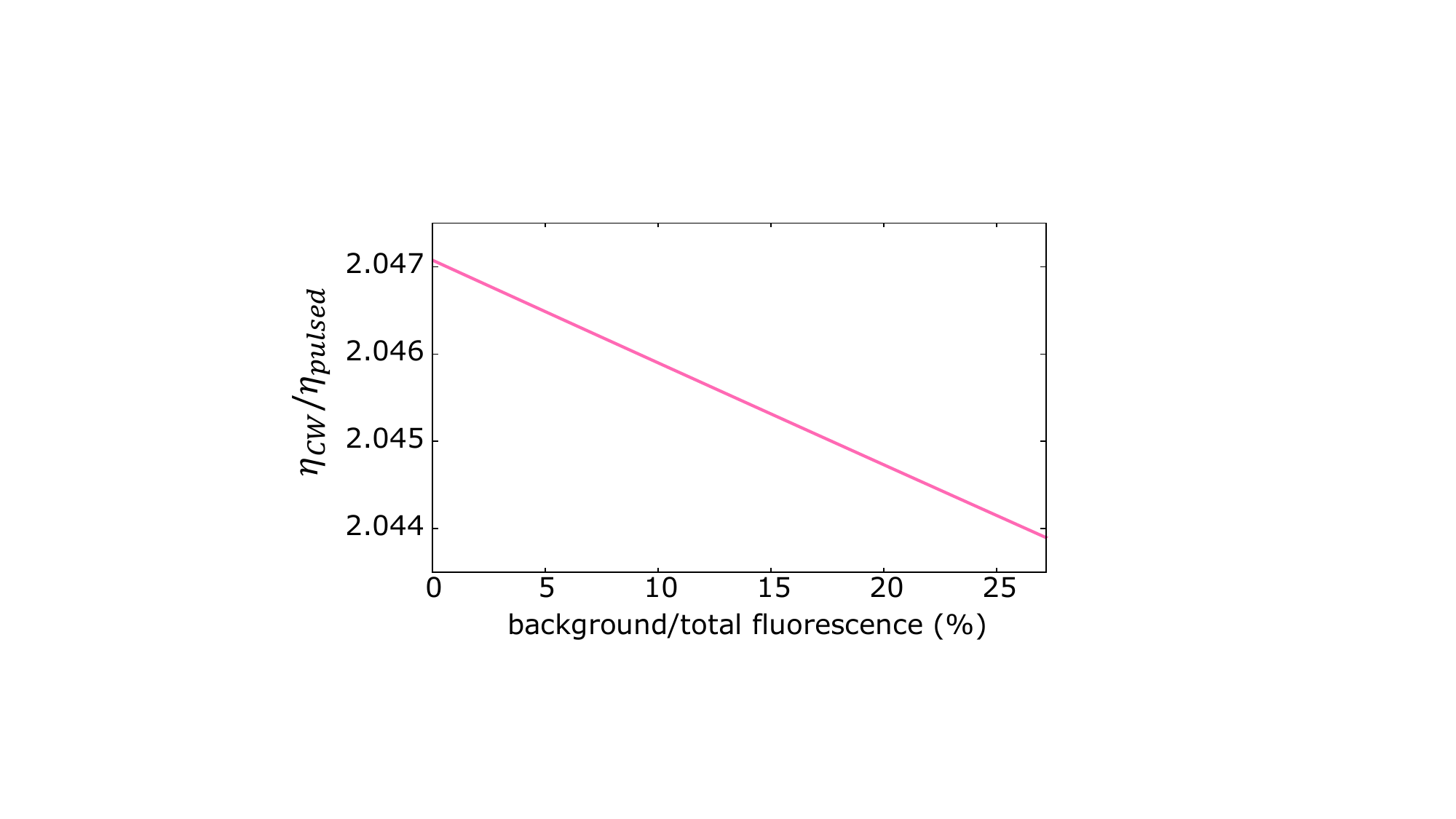}
\caption{Sensitivity ratio dependence on background. The ratio of CW to pulsed sensitivity is given for an NV ensemble ($T_2^* = 1\upmu$s, density = 300 ppb) illuminated by a beam of 100 mW with a 100 $\upmu$m waist, as a function of additional background fluorescence. At each background level, the Rabi frequency and/or pulse parameters are optimized for each protocol.} 
\label{fig:ensemblebackground}
\end{figure}

It is worth noting that the values of sensitivity we obtain may be significantly different from those  observed in experiment, for several reasons: thus far, we have not included any non-NV background fluorescence in the model, which is often significant for ensemble samples; we have neglected intrinsic spin relaxation and quasistatic dephasing in
CW ODMR; we have neglected the interplay between MW and optical pulses in deter-
mining linewidth; we have assumed perfect homogeneity in the MW drive; we have neglected strain inhomogeneity; we have made a choice for the overall collection efficiency that may not apply to a given experiment. The largest of these effects are likely background fluorescence and collection efficiency. Since the collection efficiency is a prefactor in the sensitivity, it affects the two protocols in a similar manner.  
Moreover, when we add in a background that increases linearly with optical power, we find that the CW and pulsed sensitivities are also impacted similarly. To illustrate this effect, in Fig.~\ref{fig:ensemblebackground} we consider the same diamond ($T_2^* = 1 \upmu$s, density = 300ppb) illuminated by 100 mW optical power focused to a waist of 100 $\upmu$m with an additional background fluorescence rate $\alpha R$.  For each value of $\alpha$ we find the CW and pulsed sensitivity at their respective optimal parameters (Rabi frequency and pulse timings). As $\alpha$ increases, their ratio changes little; for a background fluorescence equal to the NV fluorescence, the ratio changes by only 0.3\% relative to no background. 

\section{Hyperfine structure}
A final important consideration is the hyperfine structure of the NV center. Up to this point, our models have assumed that only a single transition is driven by microwave excitation. In most circumstances, however (away from ground- or excited-state level anticrossings), the $m_s = 0$ to $m_s = 1$ transition exhibits a triplet structure owing to hyperfine interactions with the (predominately) $I = 1$ $^{14}$N  nuclear spin. As long as the linewidth of the transition is significantly less than the hyperfine splitting (approximately $A = 2.16$ MHz~\cite{smeltzer_robust_2009}), the three transitions can be treated independently and the above analysis remains appropriate if we reduce the contrast inserted into Eq.~\ref{eq:etaCW}-\ref{eq:etapulsed} by a factor of 3. (It is tempting to infer that reducing the contrast by a factor of 3 worsens the sensitivity by the same factor of 3, but this only holds in the limit of low contrast; in general the impact on the sensitivity is greater due to the role that contrast plays in determining the shot noise at the magnetometer operation point.) 
When the hyperfine lines start to overlap, however, additional considerations arise. While a detailed investigation of the impact and mitigation of hyperfine structure is beyond the scope of this work, we can elucidate two qualitative features.

\begin{figure}[ht!]
\centering
\includegraphics[width=6cm]{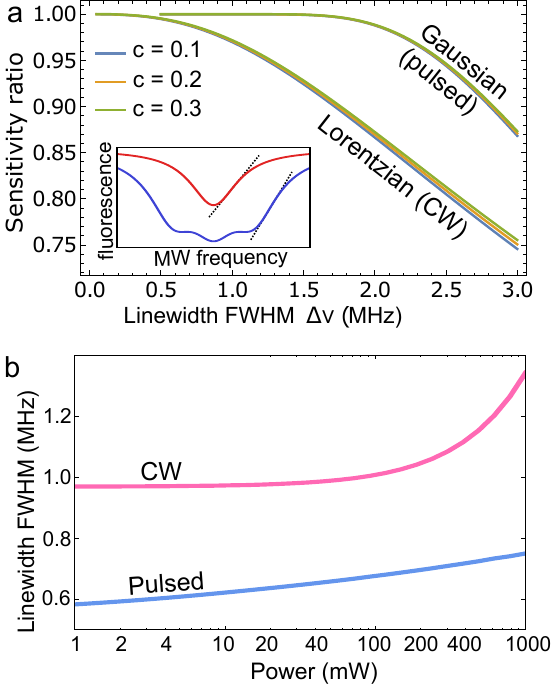}
\caption{(a) Effect of hyperfine structure on sensitivity. The plotted ratio is the sensitivity obtained at the optimal operating point for the sum of three curves (separated in frequency by $A = 2.16$MHz, each with contrast $c/3$)  divided by the sensitivity as calculated from Eq.~\ref{eq:etaCW}-\ref{eq:etapulsed} with a contrast of $c/3$. Different behaviour is observed for Gaussian (approximating pulsed ODMR) and Lorentzian (approximating CW ODMR) lineshapes. (Inset) Illustration of the mechanism for improved sensitivity showing the increase in slope associated with superposing three lines. (b) Linewidth of the CW and pulsed ODMR signals extracted from ensemble models as a function of optical power ($T_2^* = 1~\upmu$s, density = 300 ppb, beam waist = 100 $\upmu$m, as for Fig.~\ref{fig:ensemblepulsed}). }
\label{fig:hyperfine}
\end{figure}

Firstly, when driven by a single microwave tone, the overlap of the three lines tends to improve the sensitivity relative to what would be expected from a reduction in contrast by a factor of 3. Figure~\ref{fig:hyperfine}a shows the ratio of the sensitivity obtained from the superposition of three lines to what would be predicted from a threefold reduction in contrast, showing a reduction (i.e. improvement) relative to the naive expectation. This improvement arises from two effects: as illustrated in the insert to Fig.~\ref{fig:hyperfine}a, there is an increase in the slope for three superposed lines relative to one, and also a reduction in the fluorescence rate (and thus shot noise), both of which improve the sensitivity. 
The line overlap and thus the sensitivity improvement is larger for Lorentzian than Gaussian lineshapes. Furthermore, the ensemble models explored in the previous section exhibit optimal CW linewidths that generally exceed the optimal pulsed linewidth. Figure 9b illustrates optimal linewidths extracted from single-transition ensemble models with $T_2^* = 1 \upmu$s, density = 300 ppb, and beam waist = 100 $\upmu$m (as for Fig. 7). While slightly different optimal operating conditions (likely with higher Rabi frequencies) will materialize when hyperfine effects are included, the larger CW linewidths seen in single-transition models suggest that the CW protocol will benefit slightly more from the overlap of hyperfine lines than the pulsed protocol.
It is also likely that the optimal operating point for both protocols will shift to higher Rabi frequencies when hyperfine effects are included.

Secondly, to mitigate the loss of sensitivity due to hyperfine structure, many experiments drive the spin transitions by three microwave tones equally spaced in frequency by the hyperfine splitting~\cite{barry_optical_2016, ahmadi_pump-enhanced_2017, poulsen_optimal_2022}. The mitigation is only perfect in the limit of narrow lines and low Rabi frequency, such that each transition is only significantly impacted by one of the microwave tones at a time. In CW protocols, the MW Rabi frequency must compete with optical spin repolarization, and they therefore typically operate at slightly higher Rabi frequencies and larger linewidths. Consequently, pulsed protocols exhibiting narrower lines and utilizing lower Rabi frequencies are likely to benefit slightly more from the use of triple-tone excitation.

\section{Conclusion and outlook}
As NV magnetometry starts to be used in applications beyond the laboratory, practical considerations of sensor complexity, size, ease of use, power, and cost become important, restricting the choice of magnetometry protocols. Using a simple model of NV photophysics, we investigated the benefits of an increase in complexity -- moving from CW to pulsed ODMR operation -- for an ensemble device with limited optical power resources. Our results indicate a smaller benefit than might be expected from single-NV, power-unlimited experiments, illustrating the importance of considering practical constraints when comparing protocols. 

Beyond the results presented here, the four-level model we employ offers the opportunity to explore optimal operation regimes for other magnetometry techniques with limited optical power. For example, the approach taken for pulsed ODMR could be extended by replacing the $\pi$ pulse evolution with a Ramsey or dynamical decoupling sequence. By inserting the spin evolution within a photophysical model, one can include the trade-off between greater readout signal and spin polarization versus increased preparation and measurement time in the protocol optimization. 

There are also opportunities for further refinements of our model. Some of our approximations and assumptions could be revisited, for example by examining the impact of intrinsic spin relaxation, which could become important in very noisy environments and/or extremely low optical intensity, or by looking into the impact of the interplay between MW and optical pulses on ODMR linewidth. The ensemble collection efficiency could be adapted to other experimental modalities, e.g. side-collection~\cite{le_sage_efficient_2012} or parabolic concentrators~\cite{wolf_subpicotesla_2015}.  A more quantitative analysis of hyperfine structure and triple-tone excitation would add  corrections to our sensitivity comparison, and might identify slightly different optimal Rabi frequencies for operation. One could also model MW inhomogeneity to examine optimal sensing volumes. In the long run, the ability to include imperfect spin polarization and readout in modeling of power-limited devices may help to bring NV center magnetometry to a wider set of practical applications.

\section{Backmatter}

\begin{backmatter}
\bmsection{Funding}

\bmsection{Acknowledgments}
The authors thank Vincent Halde and David Roy-Guay for suggesting the problem addressed in this manuscript, and acknowledge Romain Ruhlmann, Ankita Chakravarty, Michel Pioro-Ladriere, and Khabat Heshami for productive conversations. The authors thank Vincent Jacques for allowing us to reproduce their data. 

\bmsection{Disclosures}
\medskip
LC: SB Quantum Inc. (F,R), MW: SB Quantum Inc. (R). 



\medskip



\bmsection{Data availability} Data underlying the results presented in this paper are not publicly available at this time but may be obtained from the authors upon reasonable request.

\end{backmatter}

\bibliography{magnetometry3}






\end{document}